\documentclass{ws-procs10x7}
\usepackage{balance}

\newcolumntype{d}[1]{D{.}{.}{#1}}

\makeindex
\begin{document}

\title{ANALYTIC ESTIMATE OF THE ORDER PARAMETER FOR MAGNETIC CHARGE CONDENSATION IN QCD}

\author{A. DI GIACOMO}

\address{Physics  Department, University Pisa,3  Largo B. Pontecorvo , 56127 Pisa, ITALY\\E-mail: \_adriano.digiacomo@df.unipi.it}

\twocolumn[\maketitle\abstract{The order parameter for monopole condensation is computed in terms of gauge invariant field strength correlators. Important properties emerge of the correlators in the confined phase, which could not be extracted  by existing numerical determinations on the lattice.}
]

\section{INTRODUCTION}
 A possible mechanism for confinement of color is Dual Superconductivity of the Vacuum \cite{1} \cite{2}  \cite{3}. The idea is that magnetic charges condense in the vacuum : the chromoelectric field acting between a $q-\bar q$ pair is channeled by dual Meissner effect into flux tubes  with energy
 proportional to the distance.
 
 Flux tubes are indeed observed in lattice simulations , and the corresponding field configurations are consistent with expectations \cite{3'}.
 
 An order parameter  $\langle \mu \rangle$ has been defined \cite{4} \cite{ddpp}  \cite{dp} \cite{5} \cite{6}
 which is the vacuum expectation value of a magnetically charged operator $\mu $.
  
  Numerical simulations show that  $\langle \mu \rangle \neq 0$ in the confined phase , signaling Higgs 
  breaking of magnetic $U(1)$ symmetry , $\langle \mu \rangle = 0$ in the deconfined phase. Deconfinement is an order-disorder phase transition.
  
  For  $SU(N)$ gauge group  there exist  $(N - 1)$ magnetic charges   \cite{16}and the corresponding operators  $\mu^a , (a=1, ..,N-1)$ can be written as 
  \begin{equation}
 \mu^a (\vec x, t)=e^ { {iq\over {2g}}\int d^3y \vec b_{\perp}(\vec x-\vec y)Tr(\Phi^a \vec E) (\vec y, t)}
  \end{equation}
  $q$ is an integer $\vec b_{\perp}$  is the field of a Dirac monopole with $\vec \nabla\vec b_{\perp}=0$ and $\vec \nabla\wedge \vec b_{\perp} = {\vec r\over r^3} $+ Dirac  string.
  
  $\Phi^a(x) \equiv U(x,y) \Phi^a_{d} U^{\dagger}(x,y) $ 
  
   with   $U(x,y)$ an arbitrary gauge transformation  which we shall take as a parallel transport to $x$ from a reference point $y$ along
  a path $C$ , and 
  \begin{eqnarray}
                                                   <--a -->&.&<-(N-a)>   \nonumber     \\  
  \Phi^a_{d} \equiv diag({{N-a}\over N},.,{{N-a}\over N}&,& -{a\over N},.,-{a\over N})
  \end{eqnarray}
  $\langle \mu^a \rangle$ is gauge invariant. In the gauge in which $\Phi^a =\Phi^a_d$ (Abelian Projection) 
  \begin{equation}
   \mu^a (\vec x, t)=e^ { {iq\over {2g}}\int d^3y \vec b_{\perp}(\vec x-\vec y)\vec E^a_{\perp} (\vec y, t)}
   \end{equation}
   where  $\vec E^a_{\perp}$ is the transverse 
    chromoelectric field along the color direction $T^a$
   \begin{eqnarray}
            a&.&a+1 \nonumber \\
   T^a= diag(0,...0,1&,&-1,0,..0)
   \end{eqnarray}
   and , since $\vec {E^a}_{\perp}$ is the conjugate momentum to $\vec {A^a}_{\perp}$ one has
   \begin{equation}
   \mu^a(\vec x,t) |\vec A^a_{\perp}(\vec y,t)\rangle= |\vec A^a_{\perp}(\vec y,t)+ {q\over {2g}}\vec b_{\perp}(\vec x-\vec y)\rangle
   \end{equation}
   $\langle \mu^a \rangle$ is the ratio of two partition functions \cite{dp}   with the action proportional to $\beta\equiv {{2N}\over g^2}$ . At $\beta=0$ $\mu^a = 1$. It proves convenient  to use instead of    $\langle \mu^a \rangle$ the susceptibility 
   \begin{equation}
   \rho^a \equiv {\partial \over{\partial \beta}}  ln\langle \mu^a \rangle
   \end{equation}
    in terms of which 
    \begin{equation}
    \langle \mu^a \rangle = e^{\int ^{\beta}_0 d{\beta} \rho^a(\beta)}
    \end{equation}
    $\rho^a$ is easily measured by lattice simulation, and this has been done for pure gauge theories compact $U(1)$ \cite{dp}, $SU(2)$ \cite{5} , $SU(3)$ \cite{6},  and  for $N_f=2$ $QCD$ \cite{17}   . In all cases the result is the following, in the thermodynamical limit $V\equiv L_s^3 \to \infty$
    
    1) $T < T_c$ (Confined )
    
     $\rho^a \to finite $ as $V\to \infty$  , 
     
     which by Eq(7) implies  $\langle \mu^a \rangle \neq 0$ 
    
    2)$T > T_c$ (Deconfined phase) 
    
    $\rho^a \approx -|c| L_s+c' $  or  $\langle \mu^a \rangle = 0$ 
    
    3) $T\approx T_c$ (Critical region) 
    
     ${\rho^a \over L_s^{1\over \nu}}\approx f(\tau L_s^{1\over \nu})$
    
    (finite size scaling)
    
 Here $\tau \equiv  (1-{T\over T_c} )$ ,  $\nu$ is  the critical index of the correlation length of the order parameter.
 
 $\rho^a$ is independent of the choice of the abelian projection\cite{7} \cite{8} \cite{71}.
 
 From the definition of    $\langle \mu^a \rangle $ it follows
 \begin{eqnarray}
 \langle {\mu}^a \rangle = \Sigma_0 ^{\infty} ({{iq}\over{2N}})^n {1\over {n!}} \int d^3y_1 ..d^3 y_n 
 b^{i_1}_{\perp} (\vec x -\vec y_1).. \nonumber\\ b^{i_n}_{\perp} (\vec x -\vec y_n) \langle (\Phi^a.\vec E)_{{i_1} {\vec y_1}} (\Phi^a.\vec  E)_{{i_n} {\vec y_n}} \rangle
 \end{eqnarray}
 We have used the simplified notation $(\Phi^a \vec E)_{i, \vec x} \equiv Tr [\Phi^a (t ,\vec x) E^i (t ,\vec x) ] $.

 The  $vev's$ in eq(8) are gauge invariant field strength correlators.
 We shall identify these correlators with those of the stochastic vacuum approach to QCD \cite{13a} \cite{13b} \cite{13}.
 
 \section{Cluster Expansion of $\langle \mu^a \rangle $}
 The idea of the stochastic vacuum approach is to approximate systematically the multiple gauge invariant field strength correlators by products of lowest order correlators in a systematic cluster expansion. The expansion is then truncated and only clusters up to order two are retained.
 Since the one point cluster $\langle (\Phi^a.\vec E)\rangle  = 0$, only two point clusters will survive
 the truncation, and hence only even terms in the expansion Eq(8) , which will be approximated as products of two point correlators  $\Phi^a_{i_1,i_2}(\vec y_1 -\vec y_2)= \langle (\Phi^a.\vec E)_{{i_1},{\vec y_1}} (\Phi^a.\vec  E)_{{i_2},{\vec y_2}}\rangle $.
 The combinatorial factor of the term $2n$ is $(2n - 1)!!$ and
 Eq(8) becomes
 \begin{equation}
  \langle \mu^a \rangle = e^{- {{q^2}\over {8g^2}} \int d^3 y_1 \int d^3 y_2 \Phi^a_{i_1,i_2}(\vec y_1 -\vec y_2)b^{i_1}_{\perp}(\vec y_1) b^{i_2}_{\perp}(\vec y_2)}
  \end{equation}
  or, since  $\beta = {{2N}\over{g^2}}$ Eq(6) becomes
  \begin{multline}
  \rho^a=  - {{q^2}\over {16N}}{\partial \over {\partial \beta}}[\beta \int d^3 y_1 \int d^3 y_2   \\ \Phi^a_{i_1,i_2}({\vec y}_1 -{\vec y}_2)b^{i_1}_{\perp}({\vec y}_1) b^{i_2}_{\perp}({\vec y}_2)]
  \end{multline}
  We shall identify $\Phi^a_{i_1,i_2}$ with the two point correlators defined with a straight line parallel transport which are measured on the lattice \cite{10}\cite{11}\cite{12} , and are used as an input in the stochastic model of $QCD$.  In fact for those correlators $\langle E^a E^b \rangle = \delta^{ab} \Phi$ , so that $\rho^a$ is independent on $a$ , and this agrees with lattice determinations of $\rho^a$ \cite{6}.
  
  The validity of the cluster expansion is usually justified at large distances, and we are looking for infrared properties in the study of confinement.
  
  A direct check can of be obtained by studying the dependence on the monopole charge $q$.
  The truncated $\rho$ is proportional to $q^2$ : higher correlators would introduce terms proportional to higher powers of $q$. Old data \cite{5} \cite{9'} agree with proportionality to  $q^2$ but a more systematic study of this issue 
  is planned.
  
  \section{The Field Correlators}

 There exists a general parametrization of field strength correlators dictated by invariance arguments\cite{13a} \cite{13b}
 \begin{multline}
 \Phi^{ab}_{\mu_1,\nu_1,\mu_2\nu_2}(z_1-z_2) \equiv  {1\over N}\langle Tr F^a_{\mu_1 \nu_1}(z_1)\\ V(z_1,z_2) F_{\mu_2,\nu_2}(z_2)V^{\dagger}(z_1,z_2)\rangle
 \end{multline}
 \begin{multline}
  \Phi^{ab}_{\mu_1,\nu_1,\mu_2,\nu_2}(z_1-z_2) = \delta^{ab}  \\( D(z_1-z_2)[\delta_{\mu_1\mu_2 } \delta_{\nu_1\nu_2 }-\delta_{\mu_1\nu_2 }\delta_{\nu_1\mu_2 }]   \\+{1\over 2}{\partial\over{ \partial z_{\mu_1}}}[D_1(z_1-z_2)(z_{\mu_2}\delta_{\nu_1\nu_2 } - z_{\nu_2}\delta_{\nu_1\mu_2 }]
  +  \\{1\over 2}{\partial\over{ \partial z_{\nu_1}}}[D(z_1-z_2)(z_{\nu_2}\delta_{\mu_1\mu_2 } - z_{\mu_2}\delta_{\mu_1\nu_2 }])
  \end{multline}
  At $T\neq 0$ the electric field correlators are not equal to those of the magnetic field, so one has
  four form factors, $D_E,D_{1E}, D_H,D_{1H}$ .
  For the electric field correlators $\mu_1=\mu_2=0, \nu_1=i_1,\nu_2=i_2$  and
  \begin{equation}
  \Phi^{ab}_{i_1 i_2} =\delta^{ab}[\delta_{i_1 i_2}(D_E+{1\over 2}D_{1E})+{\partial \over{\partial{z_{ i_1..}}}}]
  \end{equation}
  In the convolution with $\vec b_{\perp}$ the derivative terms give $0$. For the same reason
  the replacement can be performed 
  
  $\delta_{i_1 i_2}   \to    \delta _{i_1 i_2} - {{k_{i_1}k_{i_2}} \over {k^2}}$
  
  Going to the Fourier transform Eq(10) becomes
  \begin{multline}
  \rho^a= -{{q^2}\over 16}{\partial \over {\partial \beta}}[ \beta \int {{d^3k} \over {(2\pi)^3}} b^{i_1}_{\perp}(\vec k) b^{i_2}_{\perp}(-\vec k) \\D_E(k^2){1\over {k^2}}(k^2 \delta_{k_{i_1}k_{ i_2}}- k_{i_1} k{i_2})]
  \end{multline}
  where $\bar D_E (k^2)$ is the Fourier transform of $(D_E+{1\over 2}D_{1E})$.
  The identity
  \begin{equation}
  (k^2 \delta_{ij} -k_{i_1} k_{i_2})b^{i_1}_{\perp}(\vec k)b^{i_2}_{\perp}(-\vec k) = |\vec H(\vec k)|^2
  \end{equation} ,
  together with the explicit form of  $\vec H(\vec k)$
  \begin{equation}
  \vec H(\vec k)= \vec k \wedge \vec b_{\perp}(\vec k)
  \end{equation}
 with $\vec n$ the direction of the Dirac string ( we shall call it  $z$), gives 
  \begin{equation}
   |\vec H(\vec k)|^2 = -{1\over {k^2}}+ {1\over {k^2_z}}
   \end{equation}
   and for $\rho^a$
   \begin{multline}
   \rho^a =  {{q^2}\over 16}{\partial \over {\partial \beta}}[ \beta \int {{d^3k} \over {(2\pi)^3}}({1\over {k^2}}- {1\over {k^2_z}}) f(k^2)]
   \end{multline}
   Here $f(k^2) \equiv {1\over {k^2}}D_E (k^2)$.
   Again we notice that identifying our correlators with those of the stochastic model implies that $\rho^a$ is independent both on $a$ and on the abelian projection.
   
   At large $\beta$(deconfined phase) $f(k^2)$ can be approximated by first order perturbation
   theory 
   \begin{equation}
   f(k^2) = {1\over {2Nk}}
   \end{equation}
  the only dependence on $\beta$ is the explicit factor in Eq(18) so that
  \begin{equation}
\rho^a=  {{q^2}\over {16N}} \int {{d^3k} \over {(2\pi)^3}}{1\over k}({1\over {k^2}}- {1\over {k^2_z}}) 
\end{equation} 
The integral is easily evaluated with UV cut-off ${1\over a}$ ($a$ the lattice spacing) and IR cut-off
 ${1\over {L_s a}}$  ($L_s$ the spacial lattice size).
The result is
\begin{equation}
 \rho^a =  {{q^2}\over 16N}{1\over {(2\pi)^2}}[-\sqrt(2) L_s + 2 ln(L_s) +const]
 \end{equation}
 which, as explained in Sect. 1 means $\langle \mu^a \rangle = 0$ in the thermodynamical limit $L_s \to \infty$
 
  The two point correlators have been measured on the lattice both at $T=0$ and at finite $T$. They are well represented in the range of distances $.1fm \le  x \le 1fm$ by a form \cite{10}
  \begin{equation}
  D_E = Ae^{-{x\over \lambda_b}} + {b\over {x^4}}e^{-{x\over \lambda_a}}
  \end{equation}
  with $\lambda_a\approx 2\lambda_b$ and $\lambda_b\approx .3fm$.
  $A$ and $b$ are independent on $\beta$ from $T=0$ up to $T\approx .95 T_c$ Approaching further $T_c$ $A$ decreases rapidly to zero.\cite{12}.
  
  In fact the above parametrization cannot be valid at shorter distances, where the operator product expansion and the non-existence of condensates of dimension less than 4 require that
  \begin{equation}
  D_E \approx {b\over 2} [{1\over {(x+ie)^4} }+{1\over {(x-ie)^4} }] + c + d x^2
  \end{equation}
  The prescription on the singularity is dictated by the match to perturbation theory.
  At larger distances a stronger infrared cut-off at some distance $\Lambda$
    is expected since colored particles cannot propagate at infinite distance. This feature needs a further numerical investigation on the lattice.
  Up to  $T\approx .95 T_c$ the only dependence on $\beta$ in Eq(14) is again the explicit factor so that
  the result is the same as Eq(21) with the lattice size $L_s$ replaced by the infrared cut-off $\Lambda$
  \begin{equation}
   \rho^a =  {{q^2}\over 16}{1\over {(2\pi)^2}}[-\sqrt(2) \Lambda + 2 ln(\Lambda) +const]
   \end{equation}
   This expression gives a finite value of $\rho^a$ for $T < T_c$ and hence $\langle \mu^a \rangle \neq 0$ which means dual superconductivity for any finite value of the UV cut-off $a$.  However in the continuum limit $a \to 0$ $\rho^a$ diverges so that $\langle \mu^a \rangle $ needs a renormalization.
   This is similar to what happens for the Polyakov line in the quenched theory.\cite{biel}. Existing Lattice data support this statement [See Fig(2) of ref.\cite{5}] ,but a more systematic investigation is planned.
   When the critical temperature is approached both the IR cut-off $\Lambda$ and the coefficient $A$ in Eq(22) strongly depend on $\beta$ :$\Lambda $ diverges and $A$ tends to zero.
   A more detailed calculation, which will be reported elsewhere gives
   \begin{multline}
   \rho^a = {{q^2}\over {16N}}{\partial \over {\partial \beta}}(\beta[{1\over {(2\pi)^2}}(-\sqrt(2){\Lambda\over a}+ 2\ln({\Lambda\over a}) \\+const) +2N\lambda_b^4A_E(1+{\Lambda\over {\pi\lambda_b}})]
   \end{multline}
   Numerical determinations of the temperature dependence of $A_E$ around $T_c$ exist. Not much is known about the behavior of $\Lambda$ . Further study is needed to understand how the scaling law
   of $\rho^a$ described in Sect 1 works in Eq(25), and this  is already on the way.
   In conclusion an interesting interrelation has been found between the large distance behavior of correlators an confinement, which deserves further study.

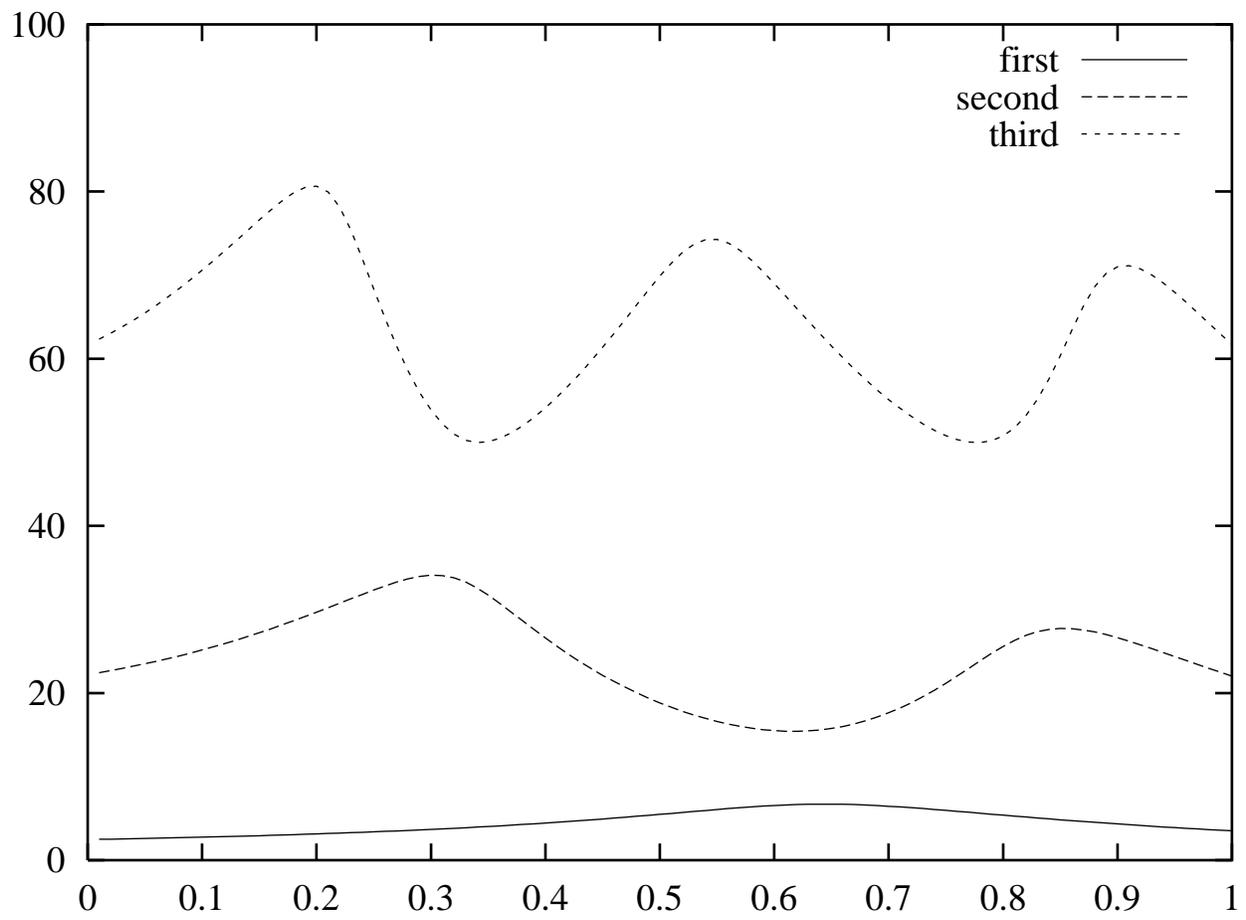
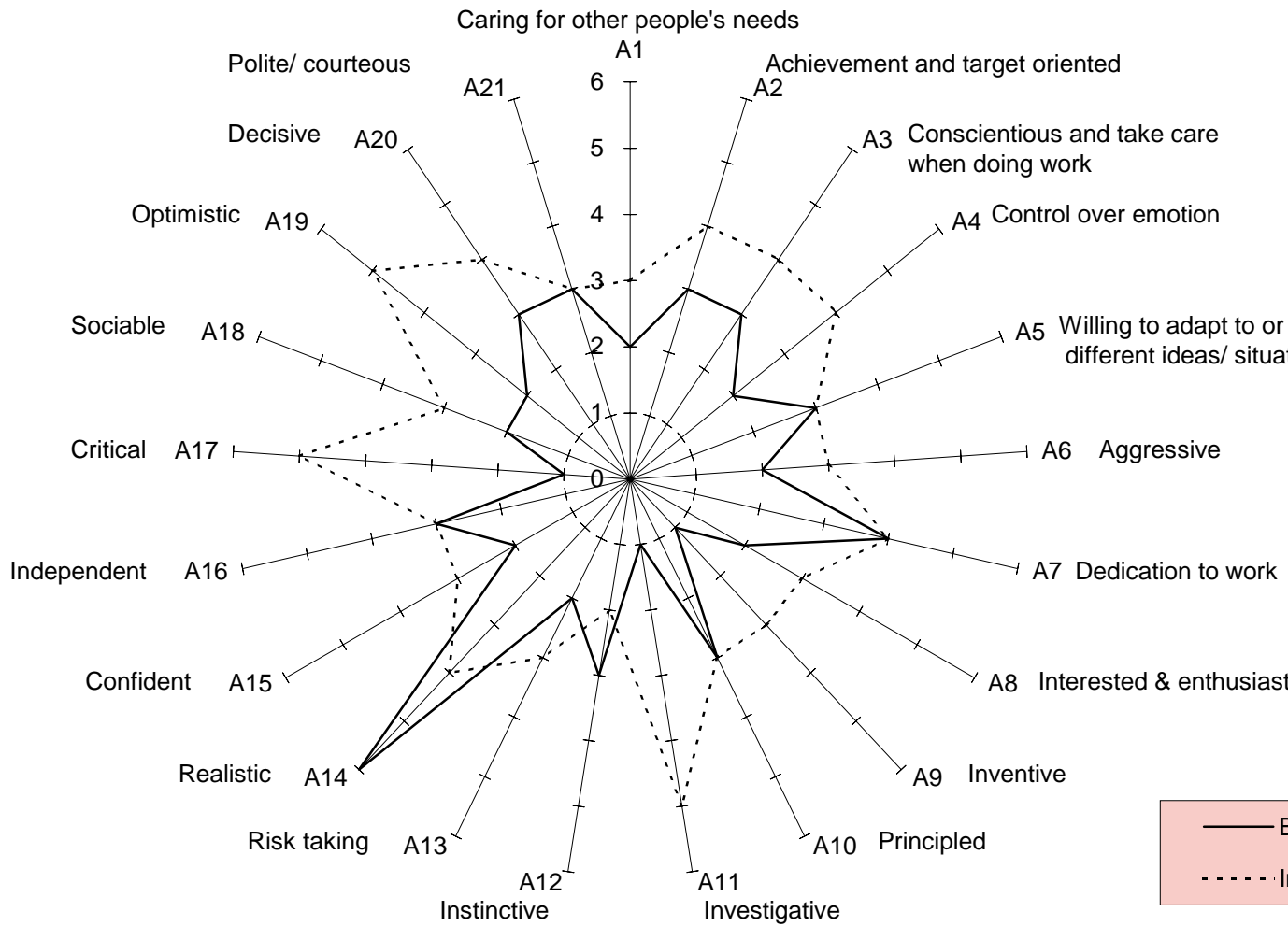
\end{document}